\documentstyle[12pt,titlepage,epsf]{article}
\def\_{\rule{.3em}{.15ex}}

\newcommand{\scs}{\scriptscriptstyle}
\def\simlt{\stackrel{<}{\scs \sim}}
   
\begin{document}
\begin{titlepage}

 \begin{flushright}
  { TPI-MINN-00/65\\
       UMN-TH-1931\\      
       hep-ph/0012123\\   
    December 2000}
 \end{flushright}

 \begin{center}
  \vspace{0.6in}

\setlength {\baselineskip}{0.3in}
  {\bf \Large The $b$--quark EDM in SUSY and CP--odd bottomonium
formation}
\vspace{2cm} \\
\setlength {\baselineskip}{0.2in}

{\large  D. A. Demir$^{^{1}}$ and
         M. B. Voloshin$^{^{1,2}}$}\\

\vspace{0.2in}
$^{^{1}}$ Theoretical Physics Institute, University of Minnesota,
Minneapolis, MN 55455

\vspace{0.2in}
$^{^{2}}$ Institute of Theoretical and Experimental Physics, Moscow,
117259

\vspace{2cm}
{\bf Abstract \\}
\end{center}
We compute the electric dipole moment (EDM) of the bottom quark in
minimal supersymmetric model (SUSY) with explicit CP violation. We
estimate its upper bound to be $10^{-20}~{\rm e-cm}$ where the dominant
contribution comes from the charginos for most of the SUSY parameter
space. We also find that chargino contribution is directly correlated
with the branching fraction of the $B\rightarrow X_s \gamma$ decay.
Furthermore, we analyze the formation of ${}^{1}\mbox{P}_{1}$ resonance
of the $(\bar{b}b)$ system in $e^+ e^-$ annihilation, and show that the
CP--violating transition amplitude, induced solely by the $b$--quark
EDM, is significantly larger than the CP--conserving ones. Therefore,
observation of this CP--odd resonance in $e^+ e^-$ annihilation 
will be a direct probe of the CP--violating phases in SUSY. In case experiment cannot
establish the existence of such a CP--odd $(\bar{b}b)$ state, 
then either sparticle masses of all three generations will be pushed 
well above TeV, weakening the possibility of weak--scale SUSY, 
or the sparticle mass spectrum will be tuned so as to cancel different 
contributions to EDMs.

\end{titlepage}
\thispagestyle{empty}
\vbox{}
\section{Introduction}
In the minimal standard model (SM) of electroweak interactions both
flavour violation and CP violation are encoded in the CKM matrix. In its
supersymmetric (SUSY) extension, however, there appear new sources for these
phenomena generated by the soft SUSY--breaking terms \cite{susyflavourCP}. In
an attempt to establish the strength and structure of the flavour and CP
violation in SUSY it is necessary to confront it with the experimental data on
flavour--changing and flavour--conserving processes. In this respect, the
flavour--conserving phenomena such as the Higgs system \cite{higgs} and the
electric dipole moments (EDM) \cite{smallphase,cancel,heavy} of particles are
useful tools in searching for new sources of CP violation in a way independent
of the flavour violation.

The existing  upper bounds on the neutron and electron EDMs \cite{expEDM}
put stringent constraints on the sources of CP violation. Even if one
solves the strong CP problem by a SUSY version \cite{strongcp} of the KSVZ
axion model \cite{pq}, the reamining electroweak contributions are to be still
suppressed. For accomplishing this, there have been several suggestions which
include ($i$) choosing \cite{smallphase} (or suppressing by a relaxation
mechanism \cite{relax}) the SUSY CP phases $\simlt {\cal{O}}(10^{-3})$, or
($ii$) finding appropriate parameter domain where different contributions
cancel \cite{cancel}, or ($iii$) making the first two generations of scalar
fermions heavy enough \cite{heavy} but keeping the soft masses of the third
generation below ${\rm TeV}$. Though each scenario
for suppressing the EDMs has its own virtues in terms of the
implied SUSY parameter space, in what follows we will work
in the framework of effective supersymmetry \cite{effsusy}
where the scenario ($iii$) can be accomodated. However,
the discussions below are general enough to be interpreted or
extended in any of the scenarios listed above.

The effective SUSY scenario deals with a single generation of
sfermions, and thus, the question of flavour--changing transitions
is avoided. Then SUSY effects can show up through the Higgs bosons,
Higgs and gauge fermions, and the third generation sfermions. In fact,
it is these light sparticles that regenerate the electron and neutron
EDMs by two--loop quantum effects \cite{edm1,edm}. Moreover, it is clear
that the third generation fermions can still have large EDMs as the
one--loop SUSY contributions cannot be suppressed for them.

In Section 2 we will compute the bottom quark EDM in effective SUSY
up to two--loop accuracy. We will see that the two--loop contributions
are directly constrained by the electron and neutron EDMs which can exist
only at two-- and higher loop levels \cite{edm}. Concerning the
one--loop effects, the chargino contribution to the bottom  EDM will be shown
to be fully constrained by the measured branching  fraction \cite{expbsgam}
of the rare $b\rightarrow X_s \gamma$ decay. On the other hand, the gluino
and neutralino contributions remain unconstrained; however, their
contributions will be seen to hardly compete with that of the charginos.

Secion 3 is devoted to a detailed discussion of the possible signatures
of a finite bottom quark EDM. In particular, we will discuss the
formation of the ${}^{1}\mbox{P}_{1}$ bottomonium level in the $e^+ e^-$
annihilation. It will be seen that the CP--violating process,
generated by the bottom EDM, dominates over the CP--conserving
ones. Therefore, possible detection of this CP--odd resonance can
be a direct probe of the bottom EDM, or equivalently, the
sources of CP--violation in SUSY.

Section 4 contains our concluding remarks.

\section{The bottom quark EDM in SUSY}
The dimension--five electric dipole operator
\begin{eqnarray}
\label{edmLag}
{\cal{L}}_{EDM}= {\cal{D}}_{b}\ \overline{b}(x)\
{\stackrel{\longleftrightarrow}{\partial_{\alpha}}}\ \gamma_5
\b(x)\ A^{\alpha}(x)
\end{eqnarray}
defines the EDM of the bottom quark at the natural mass scale of $Q\sim
m_b$. Since ${\cal{D}}_{b}$ is obtained after integrating out all heavy
degrees of freedom, it serves as a probe of the sources of CP violation at the
weak scale $Q\sim M_W$. In the SM, ${\cal{D}}_{b}$ arises at three-- and higher
loop levels \cite{sm} whereas in SUSY there exist nonvanishing contributions
already at the one--loop level \cite{smallphase}. In the SUSY parameter space
under concern, the EDM of $b$--quark receives one--loop contributions
from the exchange of gluinos (${\cal{D}}_{b}^{\tilde{g}}$), neutralinos
(${\cal{D}}_{b}^{\chi^{0}}$) and charginos 
(${\cal{D}}_{b}^{\chi^{\pm}}$). Then, including also the two--loop
contribution, the full expression for the bottom EDM reads symbolically as
\begin{eqnarray}
\label{edmb}
{{\cal{D}}_{b}}&=&{\cal{D}}_{b}^{\tilde{g}}\left[\tan\beta\sin\phi_{\mu}, 
\sin \phi_{A_b}\right]+
{\cal{D}}_{b}^{\chi^{0}}\left[\tan\beta\sin\phi_{\mu}, \sin
\phi_{A_b}\right]+
{\cal{D}}_{b}^{\chi^{\pm}}\left[\tan\beta\sin\phi_{\mu}, \sin
\phi_{A_t}\right]\nonumber\\&+&
{\cal{D}}_{b}^{2-loop}\left[\tan\beta\sin\left(\phi_{\mu} +
\phi_{A_t}\right),
\sin\left(\phi_\mu+\phi_{A_b}\right)\right]~.
\end{eqnarray}
where the dependence of the individual contributions on $\tan\beta$ and
SUSY phases is made explicit. Clearly, in the large $\tan\beta$ regime
(as large as the electron and neutron EDM bounds permit \cite{edm}), as
preferred by the recent Higgs searches at LEP \cite{lep}, the dependence of the
two--loop contribution on the sbottom sector weakens. Therefore, in this
limit ${\cal{D}}_{b}^{2-loop}$, like ${\cal{D}}_{b}^{\chi^{\pm}}$,
probes solely the stop sector whereas ${\cal{D}}_{b}^{\tilde{g}}$ and
${\cal{D}}_{b}^{\chi^{0}}$ remain sensitive to the sbottom
sector only. Moreover, in this limiting case there remains no
sensitivity to $\phi_{A_b}$ at all, and the one--loop contributions single out
$\phi_{\mu}$.

To have an estimate of the SUSY prediction for ${{\cal{D}}_{b}}$ it is
conveninent to analyze each term in (\ref{edmb}) individually. The
gluino--sbottom loop gives
\begin{eqnarray}
\label{gluana}
\left(\frac{{\cal{D}}_{b}}{e} \right)^{\tilde{g}}&=&
\left(\frac{\alpha_s(M_{SUSY})}{\alpha_s(m_t)}\right)^{16/21}\
\left(\frac{\alpha_s(m_t)}{\alpha_s(m_b)}\right)^{16/23}\  \frac{2
\alpha_s}{3 \pi}\, Q_b\, \sum_{k=1}^{2}
\Im\left[\Gamma_{\tilde{g}}^{k}\right]\, \frac{1}{M_{3}}\,
F_{0}\left(\frac{M_3^2}{M_{\tilde{b}_k}^{2}}\right)
\end{eqnarray}
where $M_{SUSY}$, representing the characteristic scale for soft masses,
is around the weak scale. The loop function $F_{0}$ as well as the vertex
mixing factors $\Gamma_{\tilde{g}}^{k}$  are defined in the Appendix.
Letting the sbottom and gluino masses be of similar order of magnitude,
one can obtain an approximate estimate of (\ref{gluana}) as
\begin{eqnarray}
\label{glu}
\left|\left(\frac{{\cal{D}}_{b}}{e} \right)^{\tilde{g}}\right| \sim 3.4\
10^{-22}~{\rm cm}\times
\left(\frac{|\mu|}{m_t}\right)\  \left(\frac{\sqrt{2}
m_t}{M_3}\right)^{3}\ \tan\beta\ \sin \phi_{\mu}
\end{eqnarray}
which can increase by one or two orders of magnitude if one streches
$\tan\beta$ up to ${\cal{O}}\left( m_t/m_b\right)$, or pushes  $|\mu|$ up to a
TeV. In making the estimate (\ref{glu}) we have assumed a relatively
heavy gluino in accord with the experimental searches \cite{pdg}.
Moreover, the GUT--type relation among the gaugino masses
$M_{3}=\frac{\alpha_s}{\alpha_2} M_2= \frac{5 \alpha_s}{3
\alpha_1} M_1$ implies that the gluino could be as heavy as a ${\rm TeV}$ if
the masses of the  lightest neutralino and chargino are to satisfy the
present bounds. In such a case the estimate given in (\ref{glu}) can be
reduced by two orders of magnitude. The predictions made here agree with those
of \cite{heavy} in that the gluino contribution may be less significant than
that of the charginos though sizes of the fine structure constants suggest the
opposite.

Next the one--loop quantum effects due to the neutralino--sbottom loops yield
\begin{eqnarray}
\label{neut}
\left(\frac{{\cal{D}}_{b}}{e} \right)^{\chi^{0}}&=&
\left(\frac{\alpha_s(M_{SUSY})}{\alpha_s(m_t)}\right)^{16/21}
\left(\frac{\alpha_s(m_t)}{\alpha_s(m_b)}\right)^{16/23}\
\frac{\alpha_1}{4 \pi}\, Q_b \sum_{k=1}^{2}
\sum_{i=1}^{4} \Im \left[\Gamma_{\chi^0}^{k i}\right]\,
\frac{1}{M_{\chi^{0}_{i}}}\,
F_{0}\left(\frac{M_{\chi^0_i}^{2}}{M_{\tilde{b}_k}^{2}}\right)
\end{eqnarray}                                
where the vertex factors $\Gamma_{\chi^0}^{k i}$ are given in the
Appendix. Using relative sizes of the fine structure constants $\alpha_s$ and
$\alpha_1$, one expects (\ref{neut}) to be roughly two orders of magnitude
smaller than the gluino contribution (\ref{glu}). Therefore, the
neutralino--induced EDM hardly competes with the gluino contribution for most
of the SUSY parameter space.

\begin{figure}[h]
\centerline{
\epsfysize = 8cm
\epsffile{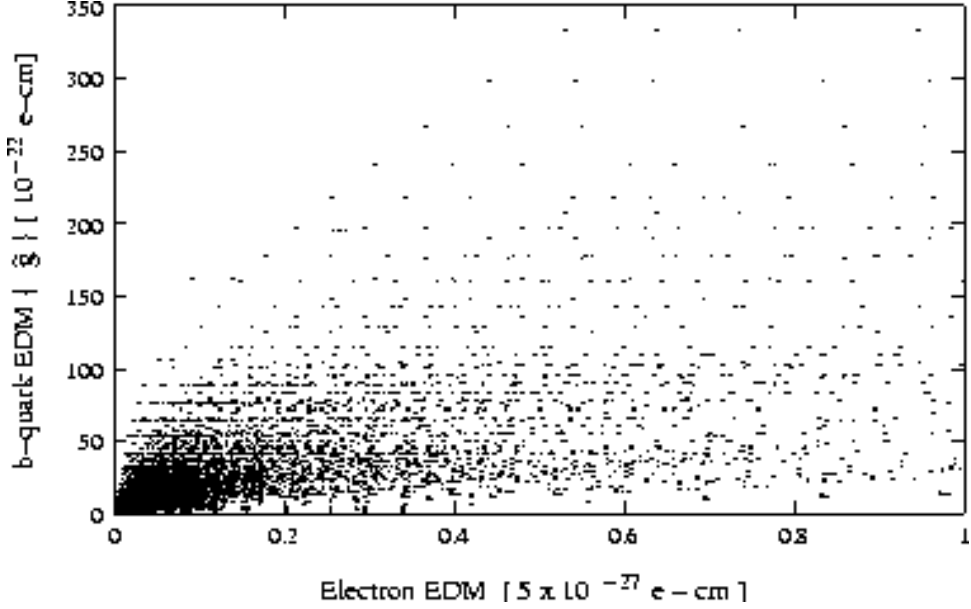}}
\caption{Variation of the gluino contribution,
$\left|{\cal{D}}_{b}^{\tilde{g}}\right|$, to the bottom EDM
(in units of $10^{-22}~{\rm e-cm}$) with the electron EDM (in units of
its present experimental upper bound
$5\times 10^{-27}~{\rm e-cm}$).}
\end{figure}                                    

Finally, the chargino--stop loop generates the last one--loop quantum
effect
\begin{eqnarray}
\label{char}
\left(\frac{{\cal{D}}_{b}}{e} \right)^{\chi^{\pm}}&=&
\left(\frac{\alpha_s(M_{SUSY})}{\alpha_s(m_t)}\right)^{16/21}
\left(\frac{\alpha_s(m_t)}{\alpha_s(m_b)}\right)^{16/23}\
\frac{\alpha_2}{4 \pi }\, Q_b\sum_{k=1}^{2}
\sum_{j=1}^{2} \Im \left[\Gamma_{\chi^{\pm}}^{k j}\right]\,
\frac{1}{M_{\chi^{\pm}_{j}}}\,
F_{\pm}\left(\frac{M_{\chi^{\pm}_j}^{2}}{M_{\tilde{t}_k}^{2}}\right)\nonumber\\
&=&- \left(\frac{\alpha_s(M_{SUSY})}{\alpha_s(m_t)}\right)^{16/21}
\left(\frac{\alpha_s(m_t)}{\alpha_s(m_b)}\right)^{16/23}\
\frac{\alpha_2}{4 \pi}\, \frac{m_b}{M_{W}^{2}}\ \Im \left[
C_{7}^{\chi^{\pm}}(M_W)\right]
\end{eqnarray}
where the first line results from the direct computation, and depends on
the vertex factors $\Gamma_{\chi^{\pm}}^{k j}$ and the loop function $F_{\pm}$
both defined in the Appendix. The second line follows from the observation that
the chargino contribution is, in fact, completely controlled by the
inclusive  $B\rightarrow X_s \gamma$ decay where $C_{7}^{\chi^{\pm}}(M_W)$ \cite{c7} 
is the Wilson coefficient associated with the electromagnetic dipole operator
${\cal{O}}_{7}=\left(e/(4 \pi)^2\right) m_b (\bar{s} \sigma^{\mu \nu}
P_R b) F_{\mu \nu}$.  The present experimental accuracy of the braching
fraction for this decay puts the bounds \cite{expbsgam}
$2.0\leq 10^{4}\times \mbox{BR}\left(B\rightarrow X_{s}
\gamma\right)\leq 4.5$ whose central value is
already consistent with the next--to--leading order SM prediction
\cite{smbsgam}. Therefore, there are rather tight constraints on the size of
the new physics contributions. For instance, it would be possible to
saturate Kaon system CP violation via pure SUSY CP phases were not it
for the $\mbox{BR}\left(B\rightarrow X_{s}
\gamma\right)$ constraint \cite{bsgamCP}. In this sense the second line
of (\ref{char}) $\left({\cal{D}}_{b}
\right)^{\chi^{\pm}}$ offers a new place where the CP violation sources
beyond the SM are constrained by the $B\rightarrow X_s \gamma$ decay. The
model--independent analyses in \cite{kagan} as well as full scanning of the
SUSY parameter space in \cite{bsgamCP2} suggest that
\begin{eqnarray}
\label{c7}
\left|\Im \left[ C_{7}^{\chi^{\pm}}(M_W)\right]\right|\simlt 1~.
\end{eqnarray}
Hence, the present experimental bounds \cite{expbsgam} imply that
\begin{eqnarray}
\label{charpred}
\left|\left(\frac{{\cal{D}}_{b}}{e } \right)^{\chi^{\pm}}\right| \simlt
2.3\times 10^{- 20}\ {\rm cm}~.
\end{eqnarray}                                                                                                  
as the characteristic size of the chargino contribution to the bottom
EDM. One notices that the bound (\ref{c7}) is valid for the entire SUSY
parameter space including $\tan\beta$ ranges as large
as ${\cal{O}}(m_t/m_b)$. This is not the case for the gluino
(\ref{gluana}) and neutralino (\ref{neut}) contributions where there is an
explicit dependence on the SUSY parameters. Furthermore, one notes
that the chargino--stop sector is under the control of the $B\rightarrow
X_s \gamma$ decay whereas the neutralino--sbottom and gluino--sbottom sectors
are largely free of direct constraints apart from collider bounds on the masses
\cite{pdg}.

\begin{figure}[h]
\centerline{
\epsfysize = 8cm
\epsffile{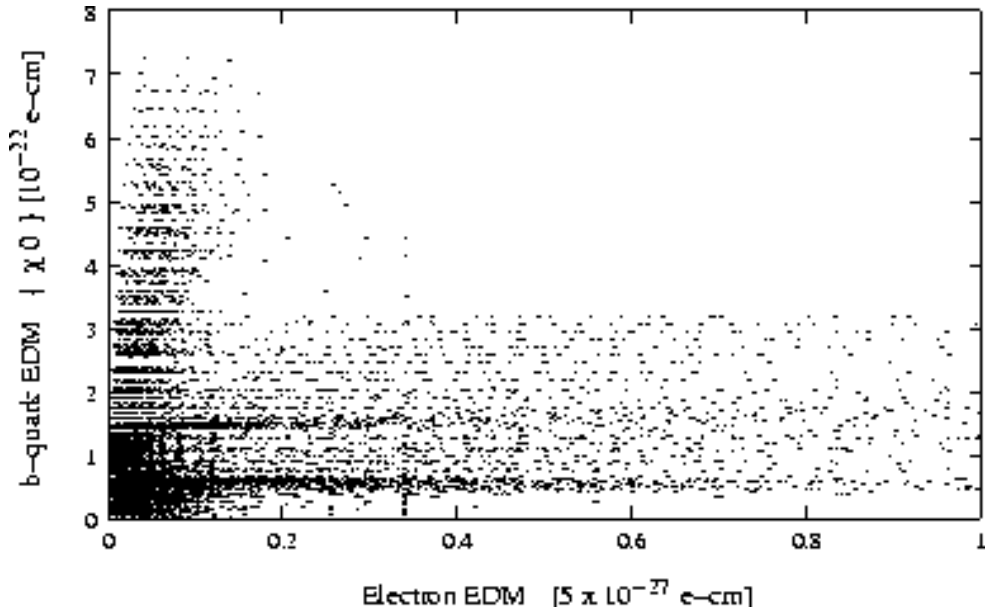}}
\caption{ The same as Fig. 1, but for neutralino contribution,
$\left|{\cal{D}}_{b}^{\chi^0}\right|$, to
the bottom quark EDM (in units of $10^{-22}~{\rm e-cm}$).}
\end{figure}        

Finally, we address the two--loop effects in (\ref{edmb}) which receive
contributions from both sbottom (decreasing with $\tan\beta$) and stop
(linearly increasing with $\tan\beta$) sectors.  It can be summarized by the
expression
\begin{eqnarray}
\label{twoloopana}
{\cal{D}}_{b}^{2-loop}= 
\left(\frac{\alpha_s(M_{SUSY})}{\alpha_s(m_t)}\right)^{16/21}
\left(\frac{\alpha_s(m_t)}{\alpha_s(m_b)}\right)^{16/23}\
\frac{m_b}{m_e}\, \left(\frac{{\cal{D}}_{e}}{e}\right)
\end{eqnarray}
where ${\cal{D}}_{e}$ is the EDM of electron which can exist only at
two--loop level \cite{edm}. The dominant contribution to ${\cal{D}}_{e}$ comes
from the pseudoscalar Higgs ($A^0$) exchange, and its present
experimental upper bound constrains the SUSY parameter space
considerably, $e.g.$, $\tan\beta\simlt 20$ for $M_{A^0}\sim m_t$. However, with
increasing $M_{A^0}$ allowed range of $\tan\beta$ expands gradually.
Then the present experimental data on ${\cal{D}}_{e}$ can be transformed
to an upper bound on the two--loop contributions to the bottom EDM using
(\ref{twoloopana}):
\begin{eqnarray}
\label{2looppred}
\left|\left(\frac{{\cal{D}}_{b}}{e } \right)^{2-loop}\right| \sim 10^{-
22}\ {\rm cm}~.
\end{eqnarray}

In the light of estimates made above, it is clear that the chargino
(\ref{charpred}) and gluino (\ref{glu}) contributions compete to dominate the
$b$--quark EDM. To check the accuracy of these approximate results, we perform
a scanning of the SUSY parameter space by varying all the mass parameters from
$m_t$ up to ${\rm TeV}$ and $\tan\beta$ from 3 to 60 in accord with
the collider bounds \cite{pdg},  recent LEP results \cite{lep}, electron
and neutron EDM upper bounds \cite{expEDM}, and the experimentally allowed
range of the $\mbox{BR}\left(B\rightarrow X_{s} \gamma\right)$ \cite{expbsgam}.

Depicted in Fig. 1 is the variation of the gluino contribution,
$\left|{\cal{D}}_{b}^{\tilde{g}}\right|$ (in units of $10^{-22}~{\rm e-cm}$),
to the bottom EDM as a function of the electron EDM (in units of the present
experimental upper bound $5\times 10^{-27}~{\rm e-cm}$). It is clear from the
figure that, ($i$) for most of the parameter space small values of the electron
EDM are prefereed, for which $\left|{\cal{D}}_{b}^{\tilde{g}}\right|
\sim 10^{-21}~{\rm e-cm}$, and and ($ii$) for certain portions of the parameter
space, where the electron EDM tends to saturate its upper bound,
$|{\cal{D}}_{b}^{\tilde{g}}$ takes on  larger values so as to dominate
the entire SUSY prediction; $\left|{\cal{D}}_{b}^{\tilde{g}}\right|_{max}\simlt
3.5\times 10^{-20}~{\rm e-cm}$. Obviously these exact results agree with the
approximate estimates made in (\ref{glu}).

Similarly, in Fig. 2 is shown the scatter plot of the neutralino
contribution, $\left|{\cal{D}}_{b}^{\chi^{0}}\right|$ (in units of
$10^{-22}~{\rm e-cm}$), as a function of the the electron EDM. It is clear
that, when the electron EDM is much smaller than the present bound,
$\left|{\cal{D}}_{b}^{\chi^{0}}\right|$ remains
mostly below $10^{-22}~{\rm e-cm}$, except for a small portion of the parameter
space where it hits in the upper bound of $10^{-21}~{\rm e-cm}$.
However, as the electron EDM takes on larger values
$\left|{\cal{D}}_{b}^{\chi^{0}}\right|$ remains bounded
around $10^{-22}~{\rm e-cm}$.

\begin{figure}[h]
\centerline{
\epsfysize = 8cm
\epsffile{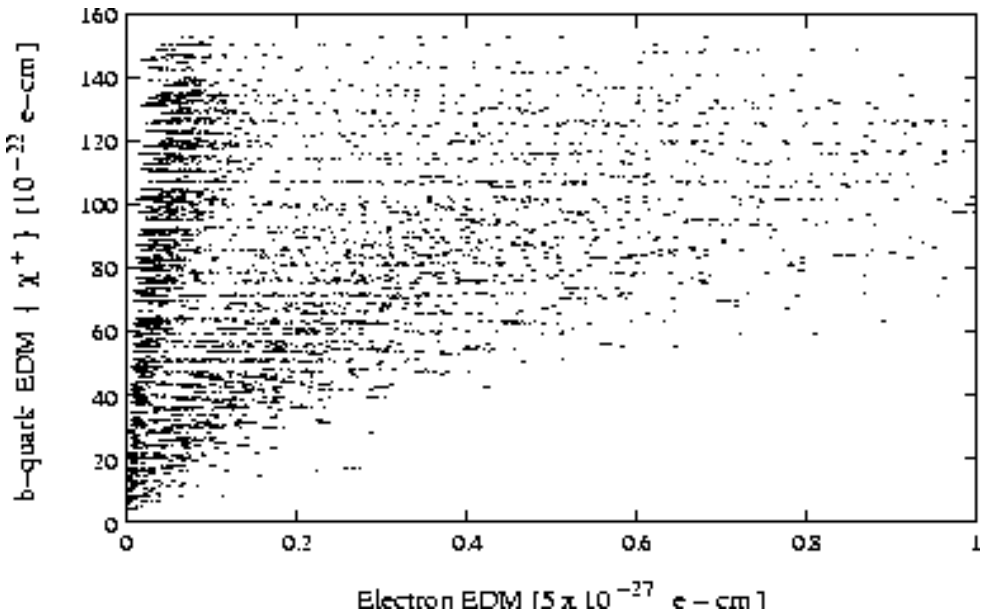}}
\caption{ The same as Fig. 1, but for the chargino contribution,
$\left|{\cal{D}}_{b}^{\chi^{\pm}}\right|$,
to the bottom quark EDM (in units of $10^{-22}~{\rm e-cm}$).}
\end{figure}                       

Fig. 3 shows is the scatter plot of the chargino contribution to the
$b$--quark EDM, $\left|{\cal{D}}_{b}^{\chi^{\pm}}\right|$) (in units of
$10^{-22}~{\rm e-cm}$) as the electron EDM varies in the experimentally allowed
range. It is clear that, for the entire range of the electron EDM,
the chargino contribution remains mostly around  $10^{-20}~{\rm e-cm}$.
That the chargino contribution, compared to the gluino one in Fig. 1, has a
shaper edge around $1.6\times 10^{-20}~{\rm e-cm}$ is a direct consequence of
the $\mbox{BR}\left(B\rightarrow X_{s} \gamma\right)$ constraint. Therefore,
Figs. 1--3 imply that ($i$) the chargino contribution is  dominant in most of
the parameter space with a value in agreement with (\ref{charpred}), ($ii$) the
gluino contribution may exceed the chargino one in certain corners of the
parameter space, ($iii$) the neutralino contribution remains of similar size as
the two--loop contribution.

As a result, the naive estimates in (\ref{glu}), (\ref{charpred}) and
(\ref{2looppred}) for different SUSY contributions to the $b$--quark EDM are
confirmed by a scanning of the SUSY parameter space as depicted in
Figs. 1--3. Consequently, in minimal SUSY the $b$--quark EDM
obeys the upper bound
\begin{eqnarray}
\label{susypred}
\left|\left(\frac{{\cal{D}}_{b}}{e} \right)\right|\simlt 10^{- 20}\ {\rm
cm}
\end{eqnarray}         
which is due to the charginos for most of the SUSY parameter space.

In principle, as long as the theory at or above the mass scale of the
fermion carries necessary sources for CP violation then the fermion
possesses an EDM. Experimentally, there is no problem in measuring the
EDM of the leptons as they can travel freely for sufficiently long distances.
For light quarks $u$, $d$ and $s$, on the other hand, EDMs make sense
due to the fact they are the constitutents of the nucleons.        

It is still meaningful to calculate the EDM of the top quark as it
can travel freely long enough distances before hadronization \cite{top}.
However, for the bottom quark the hadronization effects show up
much faster and its EDM is not observable directly. For this
reason, as in the EDMs of the $u$, $d$ and $s$ quarks, it is via the
$b$--flavoured hadrons that the bottom EDM can cause experimentally
observable  effects. Therefore, the next section is devoted to
the discussion of an experimentally testable process which is
dominated by the bottom EDM calculated above.

\section{ $b$--quark EDM and ${}^{1}\mbox{P}_{1}$ Bottomonium}

A short glance at the effective Lagrangian (\ref{edmLag}) which defines
the EDM of the $b$--quark reveals that it is, in fact, identical to the coupling 
of photon to the $ {}^{1}\mbox{P}_{1}\ (\equiv h_{b}(1 \mbox{P})$ bottomonium. 
The quantum numbers, $J^{PC}=1^{+ -}$, of this CP=-1 resonance coincide with those of
the current density \cite{charm}
\begin{eqnarray}
\label{1p1curr}
J_{\alpha}\left(\bar{b}b| {}^{1}\mbox{P}_{1}\right) =\overline{b}(x)\
{\stackrel{\longleftrightarrow}{\partial_{\alpha}}}\
\gamma_5\ b(x)
\end{eqnarray}
whose coupling to photon gives the operator structure in (\ref{edmLag}).
Presently, the experimental evidence for such CP--odd states is only
limited to the observation \cite{e760} of the charmonium $^1{\mbox P}_1$
state as a resonance in the proton-antiproton annihilation, while  the
reported signal for the bottomonium state \cite{expbottom} has
disappeared with increased statistics. In what follows, we discuss the
formation of the ${}^{1}\mbox{P}_{1}$ bottomonium in $e^+ e^-$
annihilation by an explicit calculation of the various contributions.

In the framework of the SM, $e^+ e^-$ annihilation can yield a
${}^{1}\mbox{P}_{1}$ state through the $\gamma Z$ and $Z Z$ box diagrams. The
former is the dominant process, and the relevant diagram is
shown in Fig. 4 $(a)$. The CP parities of the initial, intermediate
($\gamma \, Z$), and final states must be identical, that is,
the $e^+ e^-$ sytem has $J^{P C}=1^{+ -}$. Therefore, it is only the
longitudinal part of the $Z$ boson which contributes to the process. In other
words, the $Z$ boson exchange is equivalent to the exchange of the associated
Goldstone boson, and a straightforward calculation gives the following
effective Hamiltonian
\begin{eqnarray}
\label{smamp}
{\cal{H}}_{SM}\left({\small \mbox{CP}} \surd \right)= \frac{\alpha}{3
\pi \sqrt{2}}\; G_F 
m_e m_b\; {\cal{B}}\
J_{\alpha}\left(\bar{b}b| {}^{1}\mbox{P}_{1}\right)\cdot
J^{\alpha}\left(e^+ e^- | {}^{1}\mbox{P}_{1}\right)
\end{eqnarray}
where the current $J_{\alpha}$ is defined in (\ref{1p1curr}), and the
box function  ${\cal{B}}$ can be expressed in terms of the standard loop
integarls \cite{loop}. For the characteristic scale of the problem, 
it behaves as 
\begin{eqnarray}
{\cal{B}}\sim \frac{1}{M_Z^2\ m_b^2}\ \ln \left(\frac{m_b}{m_e}
\right)~.
\end{eqnarray}

In minimal SUSY, with two Higgs doublets, there are two CP--odd spinless
bosons, one of which becomes the longitudinal part of the $Z$ boson that
induces the effective Hamiltonian (\ref{smamp}). The other one is
the physical CP--odd Higgs scalar, $A^{0}$. Due to its CP--odd nature
this boson contributes to the formation of ${}^{1}\mbox{P}_{1}$ resonance in
$e^+ e^-$ annihilation.  Replacing the $Z$ boson  by $A^{0}$ in Fig. 4 $(a)$,
the SUSY contributions to the CP--conserving effective Hamiltonian
(\ref{smamp}) turns out to be
\begin{eqnarray}
\label{susyamp}
{\cal{H}}_{SUSY}\left({\small \mbox{CP}} \surd \right)= \tan^{2} \beta\
\times\  {\cal{H}}_{SM}\left({\small
\mbox{CP}} \surd \right) \left[ M_Z \leftrightarrow M_{A^0}\right]
\end{eqnarray}       
which rises quadratically with $\tan\beta$. If there were no constraints
coming from the electron EDM, this SUSY contribution would exceed the SM
contribution (\ref{smamp}) by three orders of magnitude for
$\tan\beta\sim 60$ and $M_{A^0} \sim M_Z$. However, it is known that
\cite{edm}, for such a light $A^{0}$, $\tan\beta\simlt 20$ so that a
conservative figure for the SUSY enhancement hardly exceeds two
orders of magnitude.

\begin{figure}[h]
\centerline{
\epsfysize = 4.5cm
\epsffile{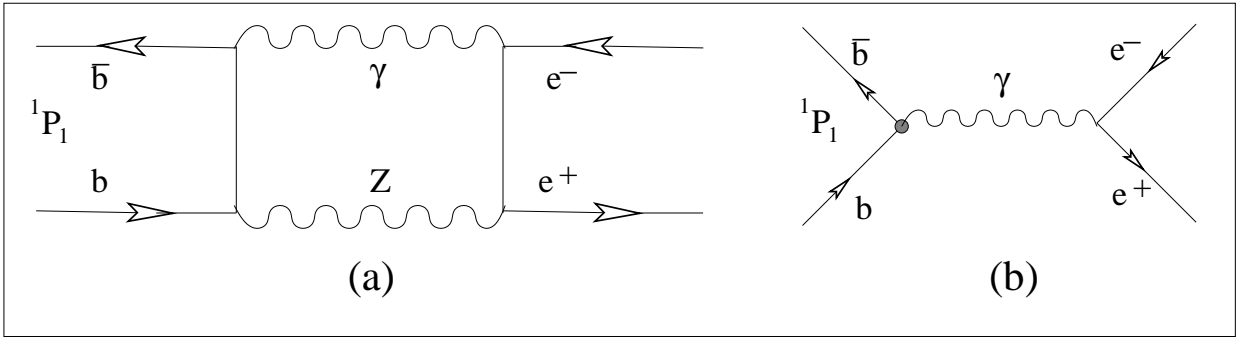}}
\caption{ Formation of the ${}^{1}\mbox{P}_{1}$ bottomonium resonance in $e^+ e^-$
scattering. The blob corresponds to the bottom quark EDM defined in (\ref{edmLag}).}
\end{figure}        

Besides the CP--conserving decay modes discussed above, the bottom quark
EDM itself can trigger the formation of ${}^{1}\mbox{P}_{1}$ state in $e^+ e^-$
annihilation. The relevant diagram is shown in Fig. 4 $(b$) where the grey blob
stands for the insertion of the effective Lagrangian (\ref{edmLag}). Due to the
CP violating nature of the EDMs it is clear that this transition violates CP so
that $e^+ e^-$ system does not need to be in the ${}^{1}\mbox{P}_{1}$ state.
In fact the effective Hamiltonian following from this diagram reads as
\begin{eqnarray}
\label{heff}
{\cal{H}}_{SUSY}\left({\small \mbox{CP}} \otimes \right)
=\left(\frac{4 \pi \alpha}{M_{h_b}^{2}}\right)\
\left(\frac{{\cal{D}}_{b}}{e} \right)\
J_{\alpha}\left(\bar{b}b| {}^{1}\mbox{P}_{1}\right)\cdot \left[e^{+}(x)\
\gamma^{\alpha}\ e^{-}(x)\right]
\end{eqnarray}
which clearly demonstrates the violation of the CP parity as the $e^+ e^-$ system is in
${}^{3}\mbox{S}_{1}$ state having CP=+1.

A comparison of the CP--conserving (\ref{smamp},\ref{susyamp}) and
CP--violating (\ref{heff}) transition amplitudes reveals that if the bottom
quark EDM falls below the critical value
\begin{eqnarray}
\label{crit}
\left|\left(\frac{{\cal{D}}_{b}}{e} \right)\right|^{crit}\sim \frac{G_F
m_e}{12 \sqrt{2} \pi^{2}}\
\frac{M_{h_b}^{2}}{M_{Z}^{2}}\ 
\ln \frac{m_b}{m_e} \ \frac{\tan^2\beta\
M_Z^2}{M_{A^0}^{2}}\sim \frac{\tan^2\beta\
M_Z^2}{M_{A^0}^{2}}\times 10^{-25}~{\rm cm}
\end{eqnarray}
then experimentally formation of ${}^{1}\mbox{P}_{1}$ bottomonium
resonance in $e^{+} e^{-}$ annihilation will not be informative at all. One
notices that this critical bound, dominated by the SUSY CP--conserving
transition (\ref{susyamp}), can be at most $10^{-23}~{\rm cm}$ which is
below (\ref{susypred}) by three orders of magnitude. This implies that the EDM
of the bottom quark is the dominant piece in forming the
${}^{1}\mbox{P}_{1}$ bottomonium in $e^{+} e^{-}$ collisions, and
observation of this resonance will become a direct probe of the soft phases in
SUSY.

The non-observation  of the ${}^{1}\mbox{P}_{1}$ state as a resonance in
$e^{+} e^{-}$ annihilation puts a model--independent bound on the bottom
EDM.  Letting $R_S (r)$ and $R_P (r)$ \cite{charm} be the
radial parts of the ${}^{3}\mbox{S}_{1}$ and ${}^{1}\mbox{P}_{1}$ levels
respectively, and using (\ref{heff}), one finds
\begin{eqnarray}
\label{bound}
\left|\frac{{\cal{D}}_{b}}{e}\right|\approx \frac{|Q_b|}{\sqrt{12}}
\left|\frac{R_S(0)}{R_{P}^{\prime} (0)}\right|
\left|\frac{\sigma\left(e^{+} e^{-} \rightarrow
{}^{1}\mbox{P}_{1}\right)}{\sigma\left(e^{+} e^{-} \rightarrow
{}^{3}\mbox{S}_{1}\right)}\right|^{1/2} \simlt 10^{-15}~{\rm cm}~.
\end{eqnarray}
The numerical value here conservatively assumes that present data
exclude the ${}^{1}\mbox{P}_{1}$ resonance in $e^{+} e^{-}$ annihilation at the
level of the formation cross section about $0.1$ of that for the
$\Upsilon$ resonance. Clearly, this result is five orders of magnitude
larger than the SUSY prediction (\ref{susypred}), and if the actual
experimental value turns out to be significantly larger than $10^{-20}$ cm,
then certainly SUSY phases will not suffice to saturate it. Especially
$\mbox{BR}\left(B\rightarrow X_{s} \gamma\right)$ will prohibit the enhancement
of the bottom EDM beyond the bounds found in the previous section.

Another way of testing the bottom EDM is in decays of the
${}^{1}\mbox{P}_{1}$, provided that a sufficiently large sample of data
for this resonance will ever be accumulated. The most direct way of searching
and testing sources of CP violation beyond the SM will be through
the decays of ${}^{1}\mbox{P}_{1}$ to hadronic final states with CP=+1.
Like the well--known $K_{L}\rightarrow \pi \pi$ decay which has established
nonvanishing CP violation in the Kaon system, decays of the form
${}^{1}\mbox{P}_{1}\rightarrow M \overline{M}$ ($M$
being a light hadron) will be a useful channel (See, for instance, \cite{pwave}
for analogous studies in charmonium system). Of course, for
the ease of experimental detection, care should be payed to choosing
appropriate final states where the CP--conserving SUSY transitions
(\ref{susyamp}) are naturally suppressed.

For instance, the decays into charmed neutral mesons,
${}^{1}\mbox{P}_{1}\rightarrow \overline{D} D$, will
proceed mainly with the  bottom EDM since the CP--conserving SUSY
contribution (\ref{susyamp}) goes like $\left(\tan\beta\right)^{0}$ as the $D$
meson side contains only up--type quarks. Moreover, for 
such a hadronic transition, the cromoelectric dipole moment (CEDM) of
the $b$--quark provides the dominant mechanism for generating an
$h_b D \bar{D}$ coupling\cite{coupling}. Although this decay mode is
preferred for enhancing the CP--violating transitions, there are various form
factors involved in the hadronic amplitude which can suppress the signal
significantly.

\section{Conclusion}
In this work we have computed the EDM of bottom quark in the minimal
SUSY model with nonvanishing soft phases. The parameter space adopted is such
that the EDMs of the neutron and electron are naturally suppressed in
that they can arise only at two and higher loop levels via the quantum
effects of scalar fermions and Higgs scalars \cite{edm}. The dominant
contribution comes from the exchange of the CP--odd Higgs scalar.

However, one notices that in the same parameter space the third
generation fermions, in particular the bottom quark, can have large EDMs
generated by the one--loop quantum effects of the scalar fermions, gluinos,
charginos and neutralinos. Indeed, in Sec. 2 we have shown, by both analytical
and numerical methods, that for most of the parameter space the chargino
contribution, which is directly correlated with the measured branching
fraction \cite{expbsgam} of the rare $B\rightarrow X_s \gamma$ decay, 
sets the upper bound on the $b$--quark EDM to be $\sim 10^{-20}~{\rm
e-cm}$. For certain corners of the parameter space the gluino contribution can
exceed this bound slightly with no order of magnitude enhancement,
however.

After estimating the $b$--quark EDM in the minimal SUSY model we have
discussed experimentally viable circumtances where it can have observable
effects. In this context, the Sec. 3 has been devoted to a detailed discussion
of the ${}^{1}\mbox{P}_{1}$ $\bar{b}b$ resonance formation in  $e^{+} e^{-}$
annihilation. The explicit calculations show that the EDM of $b$--quark is the
dominant effect in forming this CP--odd resonance, that is, the CP--conserving
transition amplitudes are below the CP--violating one by three orders of
magnitude. Hence, the very existence of a large bottom quark EDM, which is
allowed in SUSY with explicit CP violation, is the driving force behind the
possible observation of ${}^{1}\mbox{P}_{1}$ bottomonium resonance in $e^{+}
e^{-}$ annihilations. Presently the experimental bound is five
orders of magnitude above the SUSY prediction, and with increasing
precision if experiment detects such a CP--odd resonance it will be a direct
signal of the nonvanishing bottom EDM, or  equivalently, the existence
of the sources for CP violation beyond the SM such as SUSY.

However, the ultimate and most direct experimental observation of the
$b$--quark EDM will be through decays of ${}^{1}\mbox{P}_{1}$ resonance to
CP=+1 final states. In this context, one recalls the neutral charm mesons
for which the CP--conserving transition is significantly smaller than
that in the $e^{+} e^{-}$ annihilation by a factor of
$1/\tan^{2}\beta$. Therefore, especially $\overline{D} D$
type final states will prove useful in probing the strength of the $b$--quark
EDM.
                                       
If the improved experimental searches for the $h_b$ resonance in
$e^+e^-$ annihilation yield a negative result, i.e. assuming that
the present experimental precision (\ref{bound}) is improved  down to
the level of the critical value in (\ref{crit}) with no sign of
${}^{1}\mbox{P}_{1}$ resonance in $e^{+} e^{-}$ collisions, it is clear that
experiment will be no more conclusive. Even if such a resonance is observed it
will be necessary to search for its decay into CP=+1 states in order to
establish the existence of a nonvanishing $b$--quark EDM. In case
all such experimental efforts give negative results then there would
remain only two options for SUSY with nonvanishing CP phases: $(i)$ The
sparticles of all three generations are fairly above
TeV so that SUSY cannot show up at the weak scale, or $(ii)$
Contributions of various sparticle loops must cancel so as to have EDMs of
neutron, electron, muon, $b$--quark and atoms \cite{mercury} all agree
with the experimental bounds. The former makes weak scale SUSY
unlikely \cite{heavy} whereas the latter can require a finely tuned SUSY mass
spectrum \cite{cancel}.

\section*{Acknowledgements}
The work is supported in part by the US Department of Energy under the grant 
number DE-FG-02-94-ER-40823.  

\newpage
\section*{Appendix. Relevant Formulae} \setcounter{equation}{0}
\def\theequation{A.\arabic{equation}} 

\subsection*{{\large {\bf Loop Functions:}}}

The loop functions entering the evaluation of $b\rightarrow X_s \gamma$
amplitude and $b$--quark EDM are given by
\begin{eqnarray} F_{0}(a)&=&\frac{a}{2 (1-a)^2}\left[ 1+ a + \frac{2
a}{1 - a} \ln a\right]\nonumber\\
F_{\pm}(a)&=&\frac{a}{2 (1-a)^2}\left[ 7 - 5 a + \frac{2(3 - 2 a)}{1 -
a} \ln a\right]\nonumber\\
K_{1}^{8}(a)&=&\frac{1}{12 (1-a)^5}\left[ 1- 5 a - 2 a^2 - \frac{6
a^2}{1-a} \ln a \right]\nonumber\\
K_{1}^{7}(a)&=&Q_t K_{1}^{8}(a) + \frac{1}{12 (1-a)^5}\left[ 2+ 5 a -
a^2 - \frac{6 a}{1-a} \ln a \right]
\end{eqnarray}

\subsection*{{\large {\bf Mass Matrices:}}}

Here we set the conventions for the mass matrices of squarks charginos,
and neutralinos. The
mass squared matrix of the top and bottom squarks ($f=t, b$) is given by 
\begin{equation}
\widetilde{M}^2_f \ =\ \left( \begin{array}{cc} M^2_{\tilde{f}_L}\, +\,
m^2_f\, +\, \cos 2\beta M^2_Z\, ( I_{f}\,
-\, Q_f s_w^2 ) & m_f (A^*_f + \mu R_f )\\ m_f (A_f + \mu^* R_f) &
\hspace{-0.2cm} M^2_{\tilde{f}_R}\, +\,
m^2_f\, +\, \cos 2\beta M^2_Z\, Q_f s_w^2 \end{array}\right)
\end{equation}
where $R_b = R_{t}^{-1} = \tan\beta$. Being hermitian,
$\widetilde{M}^2_f$ can be diagonalized via
the unitary rotation
\begin{eqnarray}
S_{f}^{\dagger}\, \widetilde{M}^2_f\, S_{f} =
\mbox{diag.}\left(M^{2}_{\tilde{f}_1},
M^{2}_{\tilde{f}_2}\right)~,
\end{eqnarray}
with $M_{\tilde{f}_1} < M_{\tilde{f}_2}$.

The mass matrix of charginos
\begin{eqnarray} M^{-} \ =\ \left( \begin{array}{cc} M_2& - \sqrt{2} M_W
\cos \beta
\\ - \sqrt{2}M_W \sin \beta& \mu\end{array}\right)
\end{eqnarray}
can be diagnalized by a biunitary rotation
\begin{eqnarray}
C_{R}^{\dagger} M^{-} C_L = \mbox{diag.} \left(M_{\chi^{\pm}_1},
M_{\chi^{\pm}_2}\right)~,
\end{eqnarray}
where $C_R$ and $C_L$ are unitary matrices, and $M_{\chi^{\pm}_1} <
M_{\chi^{\pm}_2}$.

Finally, the neutralinos are described by a $4\times4$ mass matrix
\begin{eqnarray} M^{0} \ =\ \left(
\begin{array}{cccc} M_1& 0 & M_Z s_w \cos \beta & - M_Z s_w \sin \beta\\
0 & M_2 & - M_Z c_w \cos \beta & M_Z c_w
\sin \beta\\ M_Z s_w \cos \beta & - M_Z c_w \cos \beta & 0 & - \mu \\ -
M_Z s_w \sin \beta & M_Z c_w \sin \beta &
- \mu & 0\end{array}\right)
\end{eqnarray}
which can be diagonalized via
\begin{eqnarray}
C_{0}^{T} M^{0} C_{0} =\mbox{diag.}\left(M_{\chi^{0}_1}, \cdots,
M_{\chi^{0}_4}\right)
\end{eqnarray}                         
where $M_{\chi^{0}_1}< \cdots < M_{\chi^{0}_4}$.

\subsection*{{\large {\bf Vertex Coefficients:}}}

Here we list down the vertex coefficients entering the evaluation of the
Wilson coefficient $C_{7}$ and the
$b$--quark EDM:
\begin{eqnarray}
\Gamma_{\tilde{g}}^{k}&=&S_{{b} {1 k}}^{*} S_{{b} {2 k}}~,\nonumber\\
\Gamma_{\chi^{0}}^{k i}&=&\frac{c_w^2}{s_w^2}\ \left[ C_{{0}{2 i}}
S_{{b} {1 k}}^{*} -
 \frac{s_w}{3 c_w} C_{{0}{1 i}} S_{{b} {1 k}}^{*} - \frac{h_b}{g_2}
C_{{0}{3 i}} S_{{b} {2 k}}^{*}\right]
\left[\frac{s_w}{3 c_w} C_{{0}{1 i}} S_{{b} {2 k}} + \frac{h_b}{g_2}
C_{{0}{3 i}} S_{{b} {1
k}}\right]~,\nonumber\\
\Gamma_{\chi^{\pm}}^{k j} &=&\frac{h_b}{g_2} \left[C_{{R} {1 j}}^{*}
S_{{t} {1 k}}^{*} -
\frac{h_t}{g_2} C_{{R} {2 j}}^{*} S_{{t} {2 k}}^{*}\right] C_{{L} {2 j}}
S_{{t} {1 k}}~,
\end{eqnarray}
where the ranges of the indices are $k=1,2,\; i=1,\cdots,4$, and
$j=1,2$. In all the formulae
above, $s_w\equiv \sin \theta_w$, $c_w\equiv \cos \theta_w$ with
$\theta_w$ being the Weinberg angle.

\end{document}